\documentclass[a4paper,11pt]{article}
\pdfoutput=1 

\usepackage{jcappub} 

\usepackage[T1]{fontenc} 

\usepackage{tikz}
\usepackage{float}
\usepackage{amsmath}
\usepackage{amssymb}
\usepackage{booktabs}
\usepackage{derivative}
\usepackage{widetext} 
\usepackage{aas_macros}
\usepackage{xcolor, soul} 
\usepackage{graphicx}
\usepackage{glossaries}
\usepackage{physics}
\usepackage{siunitx}
\usepackage{tikz-3dplot}
\usepackage{subfig}
\usepackage{wrapfig}

\usepackage{hyperref}
\usepackage[capitalize]{cleveref}

\definecolor{BurntOrange}{rgb}{0.8, 0.34, 0.0}

\title{Effect of noise characterization on the detection of mHz stochastic gravitational waves}

\author[a,b]{Nikolaos Karnesis}
\author[c]{Quentin Baghi}
\author[d]{Jean-Baptiste Bayle}
\author[a,e]{Nikiforos Galanis}

\affiliation[a]{Department of Physics, Aristotle University of Thessaloniki, Thessaloniki 54124, Greece}
\affiliation[b]{Institute for Astronomy, Astrophysics, Space Applications and Remote Sensing, National Observatory of Athens, 15236 Penteli, Greece}
\affiliation[c]{Universit\'e Paris Cit\'e, CNRS, Astroparticule et Cosmologie, F-75013 Paris, France}
\affiliation[d]{IRFU, CEA, Universit\'e Paris-Saclay, F-91191 Gif-sur-Yvette, France}
\affiliation[e]{Optics Cluster Imaging Physics Department, Faculty of Applied Science, Delft University of Technology, Lorentzweg 1, 2628 CJ Delft, The Netherlands}

\emailAdd{karnesis@auth.gr}
\emailAdd{quentin.baghi@apc.in2p3.fr}
\emailAdd{j2b.bayle@gmail.com}
\emailAdd{n.galanis@student.tudelft.nl}

\abstract{
Pulsar timing arrays' hint for a stochastic gravitational-wave background (SGWB) leverages the expectations of a future detection in the millihertz band, particularly with the LISA space mission. However, finding a SGWB with a single orbiting detector is challenging: It calls for cautious modelling of instrumental noise, which is also mainly stochastic. It was shown that agnostic noise reconstruction methods provide robustness in the detection process. We build on previous work to include more realistic instrumental simulations and additional degrees of freedom in the noise inference model and analyze the impact of LISA's sensitivity to SGWBs. Particularly, we model the two main types of noise sources with separate transfer functions and power spectral density spline fitting. We assess the detectability bounds and their dependence on the flexibility of the noise model and on the prior probability, allowing us to refine previously reported results.
}

\newacronym{lisa}{LISA}{Laser Interferometer Space Antenna}
\newacronym{esa}{ESA}{European Space Agency}
\newacronym{bh}{BH}{black hole}
\newacronym{gw}{GW}{gravitational wave}
\newacronym{sgw}{SGW}{stochastic gravitational wave}
\newacronym{sgwb}{SGWB}{stochastic gravitational-wave background}
\newacronym{gr}{GR}{general relativity}
\newacronym{inrep}{INReP}{initial noise reduction pipeline}
\newacronym{tdi}{TDI}{time-delay interferometry}
\newacronym{mosa}{MOSA}{movable optical sub-assembly}
\newacronym{oms}{OMS}{optical metrology system}
\newacronym{tm}{TM}{test mass}
\newacronym[longplural={power spectral densities}]{psd}{PSD}{power spectral density}
\newacronym[longplural={amplitude spectral densities}]{asd}{ASD}{amplitude spectral density}
\newacronym[longplural={cross spectral densities}]{csd}{CSD}{cross spectral density}
\newacronym{rms}{RMS}{root mean square}
\newacronym{snr}{SNR}{signal-to-noise ratio}
\newacronym[longplural={discrete Fourier transforms}]{dft}{DFT}{discrete Fourier transform}
\newacronym{map}{MAP}{maximum a posteriori estimate}
\newacronym{dof}{DoF}{degrees of freedom}
\newacronym{mcmc}{MCMC}{Markov Chain Monte Carlo}
\newacronym{rj}{RJ}{reversible jump}
\newacronym{ss}{SS}{stepping stone}
\newacronym{pta}{PTA}{pulsar timing array}
\newacronym{ipta}{IPTA}{International Pulsar Timing Array}
\newacronym{ddpc}{DDPC}{Distributed Data Processing Center}

\tdplotsetmaincoords{60}{110}
\pgfmathsetmacro{\rvec}{1.0}
\pgfmathsetmacro{\thetavec}{30}
\pgfmathsetmacro{\phivec}{60}

\definecolor{blue}{rgb}{0.3, 0.4, 0.8}
\definecolor{amaranth}{rgb}{0.9, 0.17, 0.31}
\definecolor{pink}{rgb}{0.87, 0.56, 0.81}
\definecolor{ao}{rgb}{0.0, 0.5, 0.0}
\definecolor{maroon}{rgb}{0.76, 0.13, 0.28}
\definecolor{cardinal}{rgb}{0.77, 0.12, 0.23}
\definecolor{yellow}{rgb}{1.0, 1.0, 0.87}
\definecolor{lightseagreen}{rgb}{0.45, 0.85, 0.58}
\definecolor{gray}{rgb}{0.9, 0.9, 0.9}
\definecolor{lightblue}{rgb}{0.66, 0.84, 0.96}

\hypersetup{
	citecolor=blue,
	colorlinks=true,
	linkcolor=blue,
	urlcolor=blue
}

\DeclareSIUnit\year{yr}

\begin{document}
\maketitle
\flushbottom
\glsresetall

\section{Introduction}
\label{sec:intro}

In recent years, the ground network of \gls{gw} detectors has achieved significant advances, resulting in a large catalog of transient signals~\cite{LIGOScientific:2025hdt}. Those are mostly generated by stellar-mass \gls{bh} binaries that emit in the $\mathrm{kHz}$ range~\cite{LIGOScientific:2025slb}. One of the current and future challenges include the detection of a \gls{sgwb} signal, which might be generated by the ensemble signals of each individual astrophysical source, or may potentially have cosmological origin~\cite{LIGOScientific:2025bgj}. At the much lower frequency regime of $\mathrm{nHz}$, evidence for a stochastic gravitational background signal was announced in 2023 in a series of publications from different \gls{pta} collaborations~\cite{NANOGrav:2023gor, EPTA:2023fyk, Reardon:2023gzh, Xu:2023wog}. The future space-borne \gls{lisa} mission aims to explore the $\mathrm{mHz}$ band, bridging the gap in the \gls{gw} frequency spectrum between the \gls{pta} and ground-based detectors. 

The \gls{lisa} mission is comprised of three spacecraft forming a triangle constellation, which follows a heliocentric orbit. \gls{lisa} is scheduled to be launched in the mid-2030s. Inside each spacecraft, two test masses are hosted, being constantly held in free-falling conditions, while their relative distance between the different spacecraft is being monitored by means of laser interferometry~\cite{LISA:2024hlh}. One of the key differences of \gls{lisa} from ground-based detectors is that its interferometric arms cover millions of kilometers and use different laser sources. This requires to post-process the laser link data between the spacecraft in order to subtract the laser frequency noise. This post-processing technique is called \gls{tdi}~\cite{Tinto2002de, Estabrook2000ef, Vallisneri2004bn, pytdi}.

In this work, we start from our previous study of~\cite{Baghi:2023qnq}, relaxing several assumptions we made about instrumental noise and transfer functions. In particular, we had made the simplifying working assumption that all instrumental noises are measured via the same transfer function, which is not realistic, especially at the lower part of the frequency spectrum of the \gls{lisa} sensitivity window. In the present paper, we have dropped those assumptions by adopting more accurate individual models for the transfer functions of the various noises and signals. Similar strategies have been adopted in~\cite{Santini:2025iuj, Muratore:2023gxh}.

In addition, we have performed our analysis using a full trans-dimensional sampling when using spline models, a technique which was first introduced in the context of \gls{gw} data analysis in~\cite{Littenberg2015}. We compared the outcome of this inference to analytic template models of fixed dimension. We aim to assess the impact of the spectral model choices on the search for \gls{sgwb} in future \gls{lisa} data. The instrumental noise of the observatory will not be completely known during mission operations~\cite{LISA:2024hlh}. This is a consequence of the mission requirements, which only set upper bounds on the noise budget. Furthermore, \gls{lisa} will be a signal-dominated observatory, with multiple types of waveforms populating the data that overlap in time and frequency~\cite{LISA:2024hlh}, making it impossible to have \gls{gw}-quiet data streams. Recent works have also been focusing on the flexible spectral modeling of the noise~\cite{Santini:2025iuj, Pozzoli:2023lgz,Aimen:2025zxn}.

Taking into account the above, it is expected that any prior knowledge of instrumental noise will significantly affect our ability to discriminate the potential contributions of individual \gls{sgwb} to instrumental noise. To investigate this, we compare the performance of searches that use different levels of model flexibility for both signal and noise. To this end, we benchmark the use of parametric models versus flexible spline models for noise and signal components. We also explore the impact of the prior choice on the inference of the stochastic signal by performing the analysis under both informative and uninformative priors.

In section~\ref{sec:tf_models}, we introduce the instrument transfer functions and noise models. This analysis builds on our previous work~\cite{Baghi:2023qnq}, and arrives at a more complete transfer function model for noise and signal components. In section~\ref{sec:inference} we describe the adopted inference strategy. This includes the data pre-processing, the likelihood function, and the overall Bayesian statistical model used in the analysis. In Section~\ref {sec:results}, we present the simulations and analysis under different models and priors. Finally, in Section~\ref {sec:discussion}, we summarize and discuss our main findings.

\section{Multiple transfer function model}
\label{sec:tf_models}

\subsection{Frequency-domain data covariance}
\label{sec:total-model}

\gls{lisa}'s data are pre-processed into three \gls{tdi} channels that contain the main scientific information. In the stationary approximation, they can be discrete-Fourier transformed to get a collection of multivariate frequency series $\mathbf{\tilde{d}}(f) = (\tilde{X}(f), \tilde{Y}(f), \tilde{Z}(f))^{\intercal}$ depending on frequency $f$. The second-order statistics of these series are described by $3 \times 3$ spectrum matrices $\mathbf{C}_{d}(f)$ which include the contribution of all stochastic processes in the data. In this work, we assume that they have two origins: either instrumental noise sources or \gls{sgwb}. We denote by $\mathbf{C}_{d, n}(f)$ the noise contribution and $\mathbf{C}_{d, \mathrm{GW}}(f)$ the \gls{sgwb} contribution to the total covariance, so that
\begin{equation}
\label{eq:data-covariance}
    \mathbf{C}_{d}(f) = \mathbf{C}_{d, \mathrm{GW}}(f) + \mathbf{C}_{d, n}(f).
\end{equation}

We showed that for an isotropic background, the \gls{psd} can be factored out as
\begin{equation}
\label{eq:gw-covariance}
    \mathbf{C}_{d, \mathrm{GW}}(f) = S_{h}(f, \boldsymbol{\theta}_{\mathrm{GW}}) \mathbf{R}_{\mathrm{GW}}(f, t_0),
\end{equation}
where $S_{h, \boldsymbol{\theta}_{\mathrm{GW}}}(f)$ is the \gls{sgwb}'s strain \gls{psd} and $\mathbf{R}_{\mathrm{GW}}(f, t_0)$ is a $3 \times 3$ matrix encoding the data correlations from the \gls{sgwb}. In this work, we ignore the dependence on $t_0$ due to the relatively small variations in arm length for the chosen orbits. The strain \gls{psd} is described by a template depending on a parameter vector $\boldsymbol{\theta}_{\mathrm{GW}}$. 

We extend the noise model presented in Paper I by considering two different noise sources: the high-frequency \gls{oms} noise and the low-frequency \gls{tm} noise, each with its own response matrix. If we assume that each noise source is described by an identical \gls{psd} for all interferometers and that the two noise types are completely independent, the noise covariance in \cref{eq:data-covariance} can be decomposed as 
\begin{equation}
\label{eq:noise-covariance}
    \mathbf{C}_{d, n}(f) = S_{\mathrm{OMS}}(f, \boldsymbol{\theta}_{\mathrm{OMS}}) \mathbf{R}_{\mathrm{OMS}}(f, t_0) + S_{\mathrm{TM}}(f, \boldsymbol{\theta}_{\mathrm{TM}}) \mathbf{R}_{\mathrm{TM}}(f, t_0),
\end{equation}
where $S_{\mathrm{\alpha}, \boldsymbol{\theta}_{\alpha}}(f)$ and $\mathbf{R}_{\alpha}(f, t_0)$ are respectively the \gls{psd} and response matrix of noise $\alpha$. Each \gls{psd} is parametrized by a vector $\boldsymbol{\theta}_{\alpha}$. For the analysis, we usually define the concatenated noise parameter vector $\boldsymbol{\theta}_{n} \equiv \cup_{\alpha} \boldsymbol{\theta}_{\alpha}$. Note that the response matrix encodes all correlations due to \gls{tdi} processing, and is assumed to be perfectly known.

In this modeling, the parameters $\boldsymbol{\theta}_{\mathrm{GW}}$ and $\boldsymbol{\theta}_{n}$ control the shape of the component source \glspl{psd} $S_{\alpha}(f)$, while the response matrices $\mathbf{R}_{\alpha}$ remain fixed. This approach has two advantages. First, it avoids fitting the dips in the \gls{tdi} spectrum, which are due to the combination of the delayed interferometry measurements (causing destructive interference in the transfer function; see, e.g.,~\cite{Muratore:2022nbh}). Second, the spectral shapes of the component \glspl{psd} $S_{\alpha}(f)$ are smoother than the elements of the covariance $\mathbf{C}_d(f)$, allowing for a reduced number of spline components in shape-agnostic models (see later sections~\ref{sec:spline-models}).

\subsection{Noise responses}
\label{sec:noise_responses}

In this section, we give the form of the noise component response matrices $\mathbf{R}_{\alpha}(f,t_0)$. They relate to transformation matrices $\mathbf{M}_{\alpha}(f, t_0)$ as 
\begin{equation}
    \mathbf{R}_{\alpha}(f, t_0) = \mathbf{M}_{\alpha}(f, t_0) \mathbf{M}_{\alpha}(f, t_0)^{\dagger},
\end{equation}
where $\mathbf{M}_{\alpha}(f, t_0)$ have size $3 \times 6$  and allow for the transformation of a given interferometric noise $\mathbf{\tilde{n}}_{\alpha}(f)$ (expressed in frequency domain) into \gls{tdi} variables:
\begin{equation}
    \mathbf{\tilde{d}}_{\alpha}(f) = \mathbf{M}_{\alpha}(f, t_0) \mathbf{\tilde{n}}_{\alpha}(f),
\end{equation}
where $\mathbf{\tilde{d}}_{\alpha}(f)$ is the contribution of component $\alpha$ to the \gls{tdi} data. The matrix $\mathbf{M}_{\alpha}(f, t_0)$ encodes the full \gls{tdi} processing, including the formation of the intermediary variables $\mathbf{\eta}$ to suppress spacecraft jitter noise and reduce the effective number of lasers (see \cite{quang_nam_time-delay_2023}). This step can be performed through a $6 \times 6$ matrix $\mathbf{M}_{\alpha, n \rightarrow \eta}(f, t_0)$. Then, the Michelson \gls{tdi} variables can be constructed from the intermediary variables through another matrix of size $3 \times 6$ that we call $\mathbf{M}_{\eta \rightarrow d}(f, t_0)$, so that the full transformation matrix is
\begin{equation}
    \mathbf{M}_{\alpha}(f, t_0) = \mathbf{M}_{\eta \rightarrow d}(f, t_0) \mathbf{M}_{\alpha, n \rightarrow \eta}(f, t_0).
\end{equation}
Note that $\mathbf{M}_{\eta \rightarrow d}(f, t_0)$ is common to all noise sources $\alpha$, as the variables $\eta$ are always transformed the same way for a given \gls{tdi} combination. However, $\mathbf{M}_{\alpha, n \rightarrow \eta}(f, t_0)$ may differ depending on the noise source.

We start by considering the readout noise in the science interferometric measurement (SCI), which we assimilate in this study to \gls{oms} noise. The transformation matrix to form the variables $\eta$ is simply
\begin{equation}
    \mathbf{M}_{\mathrm{OMS}, n \rightarrow \eta} = \operatorname{diag}\left(\theta_{12}^{\mathrm{sci}}, \theta_{23}^{\mathrm{sci}}, \theta_{31}^{\mathrm{sci}}, \theta_{13}^{\mathrm{sci}}, \theta_{32}^{\mathrm{sci}}, \theta_{21}^{\mathrm{sci}}\right),
\end{equation}
where $\theta_{ij}^{\mathrm{sci}}$ are the beatnote polarizations of the SCI interferometer. In this study, we assume all polarizations are equal to unity, so that $\mathbf{M}_{\alpha, n \rightarrow \eta}$ is equal to the identity matrix $\mathbf{I}_{6}$. The first index $i$ in the indexing $ij$ refers to the spacecraft index of the considered \gls{mosa}, while the second index refers to the distant \gls{mosa} we point at. This matrix definition assumes a particular (arbitrary) ordering for the interferometric measurements in the vector $\mathbf{n}_{\mathrm{OMS}}$, with $ij = 12,\, 23,\, 31,\, 13,\, 32,\, 21$.

Regarding the \gls{tm} acceleration noise present in the \gls{tm} interferometer, assuming no correlations within spacecraft, its components appear twice in each variable $\eta_{ij}$, as it involves the subtraction of the \gls{tm} interferometer $ij$ and $ji$, so that 
\begin{equation}
    \eta_{ij, \mathrm{TM}} = - \mathbf{D}_{ij} n_{ji, \mathrm{TM}} - n_{ij, \mathrm{TM}},
\end{equation}
where $n_{ij,\mathrm{TM}}$ is the projection of the \gls{tm} jitter onto the sensitive axis and $\mathbf{D}_{ij}$ is the delay operator, which applies a delay $L_{ij}(t)$ equal to the light travel time from spacecraft $j$ to $i$. In the frequency domain, such an operator can be approximated by a complex phasor:
\begin{equation}
    \tilde{\mathbf{D}}_{ij}(f, t_0) = e^{-2 i \pi f L_{ij}(t_0)}.
\end{equation}
This approximation holds if the delay variations in time can be neglected, for example, when considering orbits with constant arms, or when considering durations short enough so that the approximation holds. In this work, the former case applies, as we use Keplerian orbits at first order in eccentricity.
With our choice of ordering, we have
\begin{equation}
    \mathbf{M}_{\mathrm{TM}, n \rightarrow \eta} = \begin{pmatrix} 
    -1 & 0 & 0 & 0 & 0 & -\tilde{\mathbf{D}}_{12} \\
    0 & -1 & 0 & 0 & -\tilde{\mathbf{D}}_{23} & 0 \\
    0 & 0 & -1 & -\tilde{\mathbf{D}}_{31} & 0 & 0 \\
    0 & 0 & -\tilde{\mathbf{D}}_{13} & -1 & 0 & 0 \\
    0 & -\tilde{\mathbf{D}}_{32} & 0 & 0 & -1 & 0 \\
    -\tilde{\mathbf{D}}_{21} & 0 & 0 & 0 & 0 & -1\end{pmatrix},
\end{equation}
where we dropped the dependence on frequency and time to simplify the notation.

Finally, the common transformation matrix $\mathbf{M}_{\eta \rightarrow d}$ is completely determined by the definition of second-generation Michelson \gls{tdi} variables~\cite{bayle_effect_2019,nam_tdi_2023}:
\begin{align}
\label{eq:tdi_x2}
X_2= & \left(1-\mathbf{D}_{121}-\mathbf{D}_{12131}+\mathbf{D}_{1312121}\right)\left(\eta_{13}+\mathbf{D}_{13} \eta_{31}\right) \nonumber \\
& -\left(1-\mathbf{D}_{131}-\mathbf{D}_{13121}+\mathbf{D}_{1213131}\right)\left(\eta_{12}+\mathbf{D}_{12} \eta_{21}\right)
\end{align}
where we used the short-hand convention $\mathbf{D}_{i_1 \dots i_n} = \Pi_{j=1}^{n-1} \mathbf{D}_{i_{j}i_{j+1}}$. For convenience we rewrite eq.~\eqref{eq:tdi_x2} as
\begin{equation}
    X_2 = \mathbf{A}_{13}\left(\eta_{13}+\mathbf{D}_{13} \eta_{31}\right) - \mathbf{A}_{12}\left(\eta_{12}+\mathbf{D}_{12} \eta_{21}\right),
\end{equation}
where we defined the operators $\mathbf{A}_{13} = 1-\mathbf{D}_{121}-\mathbf{D}_{12131}+\mathbf{D}_{1312121}$ and $\mathbf{A}_{12} = 1-\mathbf{D}_{131}-\mathbf{D}_{13121}+\mathbf{D}_{1213131}$.

We compute the variables $Y_2$ and $Z_2$ by cyclic permutations of indices ($1 \rightarrow 2 \rightarrow 3 \rightarrow 1)$. One can rewrite the full \gls{tdi} transformation in the frequency domain, and express it as a matrix $\mathbf{M}_{\eta \rightarrow d}$:
\begin{eqnarray}
\label{eq:tdi_matrix}
    \mathbf{M}_{\eta \rightarrow d} = \begin{pmatrix}
        -\tilde{\mathbf{A}}_{12} & 0  &  \tilde{\mathbf{A}}_{13} \tilde{\mathbf{D}}_{13} & \tilde{\mathbf{A}}_{13} & 0 & -\tilde{\mathbf{A}}_{12}  \tilde{\mathbf{D}}_{12} \\
        \tilde{\mathbf{A}}_{21} \tilde{\mathbf{D}}_{21} & -\tilde{\mathbf{A}}_{23} & 0 & 0 & -\tilde{\mathbf{A}}_{23}  \tilde{\mathbf{D}}_{23} &  \tilde{\mathbf{A}}_{21} \\
        0 &  \tilde{\mathbf{A}}_{32} \tilde{\mathbf{D}}_{32} & -\tilde{\mathbf{A}}_{31} & -\tilde{\mathbf{A}}_{31} \tilde{\mathbf{D}}_{31}&  \tilde{\mathbf{A}}_{32}& 0 
    \end{pmatrix}.
\end{eqnarray}

\subsection{Gravitational wave response}
\label{sec:gw_responses}

Similarly to the noise, we compute the \gls{gw} response in two steps. First, we derive the single-link (or $\eta$) response, and second, we apply the \gls{tdi} transformation.
The computation can be decomposed as
\begin{equation}
    \mathbf{R}_{\mathrm{GW}}(f, t_0) = \mathbf{M}_{\eta \rightarrow d}(f, t_0) \left( \mathbf{R}_{+}(f, t_0) + \mathbf{R}_{\times}(f, t_0)\right)\mathbf{M}_{\eta \rightarrow d}(f, t_0) ^{\dagger},
\end{equation}
where $\mathbf{M}_{\eta \rightarrow d}$ is the \gls{tdi} transformation matrix given by eq.~\ref{eq:tdi_matrix} and $\mathbf{R}_{+},\mathbf{R}_{\times}$ are the \gls{sgwb} single-link response matrices for the two strain polarizations.
For a given polarization $p$, the response matrix is computed numerically assuming $N$ stochastic point sources evenly distributed in the sky sphere, so that we have schematically
\begin{equation}
    \mathbf{R}_{p}(f, t_0) = \sum_{i=1}^{N} \mathbf{G}_{p}(f, t_0, \mathbf{\hat{k}}_i) \mathbf{G}_{p}(f, t_0, \mathbf{\hat{k}}_i)^{\dagger},
\end{equation}
where $\mathbf{G}_{p}$ is a response kernel matrix depending on source sky location $\mathbf{\hat{k}}$. The expression for its elements are given in Appendix B of Paper I~\cite{Baghi:2023qnq}.

\section{Stochastic Gravitational Wave Background inference}
\label{sec:inference}

\subsection{Data pre-processing}

We define the multivariate \glspl{dft} $\mathbf{\tilde{d}}(f_k)$ as in~\cite{Baghi:2023qnq} and the periodogram matrix as
\begin{eqnarray}
    \mathbf{P}(f_{k}) = \mathbf{\tilde{d}}(f_k) \mathbf{\tilde{d}}(f_k)^{\dagger}.
\end{eqnarray}
The periodogram matrix is an estimator of the full covariance matrix $\mathbf{C}_d(f_k)$. Its expectation converges to it in the limit of infinite time series.
To reduce computational cost, we restrict the analysis to a bandwidth with boundaries $f_{\min}$ and $f_{\max}$, and split it into segments whose sizes increase with frequency. From each segment labeled $j$, we build an average of the periodogram matrix as
\begin{equation}
	\label{eq:averaged-periodogram}
    \mathbf{\bar{P}}(f_j) \equiv \frac{1}{2K_j+1} \sum_{k=j-K_j}^{j+K_j} \mathbf{P}(f_k),
\end{equation}
where $K_j$ is the number of frequency bins around $f_j$, so that $n_j = 2 K_j +1$ is the number of frequency bins averaged in segment $j$.

\subsection{The likelihood model}
\label{sec:likelihood}

The average periodogram defined in \cref{eq:averaged-periodogram} would follow a complex Wishart distribution with $n_j$ degrees of freedom if the \glspl{dft} bins $\mathbf{\tilde{d}}(f_k)$ were not correlated. However, due to windowing, frequencies are not uncorrelated, thus we need to approximate the distribution using a Wishart distribution with an effective number of degrees of freedom $\nu(f_j)$ that is smaller than the number of averaged periodogram bins $n_j$ and a scale matrix $\mathbf{C}_{d}(f)$. After dropping the constant terms, the log-likelihood writes
\begin{equation}
	\mathcal{L}(\boldsymbol{\theta} )= - \sum_{j=j_{\min}}^{j_{\max}} \nu(f_j) \left[ \operatorname {tr} (\mathbf {C}_{d}^{-1}(f_j) \mathbf{\bar{P}}(f_j)) + \log |\mathbf {C}_{d}(f_j)| \right].
	\label{eq:log-likelihood}
\end{equation}
The appropriate choice for $\nu(f_j)$ depends mainly on the size of the segment centered in $f_j$ and on the window used to taper the time series. We provide details on how to compute it in Ref.~\cite{paper-in-prep}. In this work here we have adopted a conservative figure of $\nu(f_j) = n_j / N_\mathrm{bw}$, where $N_\mathrm{bw}$ is the normalized equivalent noise bandwidth, which depends on the given window function~\cite{Heinzel2002SpectrumAS}. 

\subsection{Model selection}
\label{sec:modelselect}

In a Bayesian framework, detecting the presence of a stochastic process can be done through model comparison: one model assumes that the data only contains noise (null hypothesis $H_0$), while the other model assumes the presence of a \gls{sgwb} in addition to the noise (tested hypothesis $H_1$). We compare the models by computing their Bayes factor, defined as the ratio of their evidences. The log-Bayes factor is then
\begin{equation}
	\log \mathcal{B}_{10} = \log Z_{1} - \log Z_{0},
	\label{eq:bayes_fact}
\end{equation}
where $Z_i \equiv \int_{\Theta_{i}} p\left(\mathbf{\bar{P}}\vert  \boldsymbol{\theta}_i, H_i\right) \dd \boldsymbol{\theta_i}$ is the evidence of the model under hypothesis $H_i$ and $\Theta_{i}$ is the space in which $\boldsymbol{\theta}_{i}$ is allowed to take values. For example, the presence of a \gls{sgwb} is claimed when the Bayes factor exceeds a given threshold. The appropriate threshold value is usually a challenging quantity to compute. However, it can be chosen following recommendations in the literature (e.g., from~\cite{Kass:1995loi}), or evaluated by characterizing the Bayes factor distribution from many data realizations, as done in previous works~\cite{Baghi:2023qnq, Adams2010vc}. Other studies can also be utilized in this regard, such as~\cite{Pozzoli:2024hkt, Karnesis2019mph}. 

Running multiple tempered \gls{mcmc} chains in parallel allows us to use different methods that can approximate the log-evidence $\log Z_i$ of a given model. The first is the  thermodynamic integration~\cite{lartillot_computing_2006},
\begin{equation}
	\log Z_{i} =\int_{0}^{1} \mathrm{E}_{\beta}\qty[\log p\left(\mathbf{\bar{P}} \vert \boldsymbol{\theta}_i, H_{i} \right)] \dd{\beta},
\end{equation}
where the expectation $\mathrm{E}_{\beta}$ is taken with respect to the tempered posterior density $p_{\beta}\left(\mathbf{\bar{P}}\vert \boldsymbol{\theta}_i, H_{i} \right) \propto  p\left(\mathbf{\bar{P}} \vert \boldsymbol{\theta}_i, H_{i} \right)^\beta p(\boldsymbol{\theta}_i, H_i)$. The variable $\beta$ is the inverse temperature of the tempered chain, and $\mathrm{E}_{\beta}$ is the expectation of the chain at temperature $1/\beta$ taken over the parameter space $\Theta_i$. The second is the so-called \gls{ss} algorithm~\cite{ss1, 10.1093/mnras/staf953}, which uses random segments of the \gls{mcmc} chains to compute the evidence as
\begin{equation}
\label{eq:ss}
    \log Z_i = \sum_{k=1}^{N_T-1}\log \sum_{i=1}^n \log p\left(\mathbf{\bar{P}} \vert \boldsymbol{\theta}_i, H_{i} \right)^{\beta_{k+1} - \beta_{k}} - (N_T - 1) \log n ,
\end{equation}
where $N_T$ is the number of temperatures used in the particular run, and $n$ the number of posterior samples used in the calculation. This method is proven to be more robust in cases where the number of temperatures is relatively small~\cite{ss2}. Recently there have been more methods introduced in the literature, such as the one of~\cite{Zahraoui:2025zte}, but we choose to work with the \gls{ss} algorithm, due to its availability within the Eryn package. 

When the number of competing models is large, the evidence computation becomes computationally prohibitive because the result of the inference is necessary with all candidate models available. The solution to such a problem would be to sample for the optimal model order simultaneously with the model's parameters. This is achievable by employing \gls{rj} \gls{mcmc} algorithms~\cite{green:1995}. The \gls{rj} part refers to having the ability to make proposals, or ``jumps'', between models with different dimensionality. The \gls{rj} \gls{mcmc} can be seen as a generalization of the commonly used Metropolis-Hastings algorithm~\cite{green:1995}. For this work we use {\tt Eryn}~\cite{Karnesis:2023ras}, which is an affine-invariant sampling tool~\cite{emcee}, which brings together \gls{rj}\gls{mcmc} and parallel tempering techniques~\cite{Vousden2016}. 

Since we are interested in defining detectability bounds for signals with different characteristics, we use the Stepping Stone method to estimate the log-evidences of the signal plus noise $Z_{1}$, and noise-only $Z_{0}$ models, respectively. Then a comparison between the two models is given directly by computing the Bayes Factor from eq.~(\ref{eq:bayes_fact}). On the other hand, as we shall see in section~\ref{sec:spline-models}, we fit flexible spectral shapes based on spline interpolation, where the optimal number of spline knots is estimated via \gls{rj} \gls{mcmc} sampling algorithms.

\subsection{Shape-agnostic spectral models}
\label{sec:spline-models}

Following our previous work~\cite{Baghi:2023qnq}, we can adopt a flexible model for a given \gls{psd}. This flexible model is based on spline-interpolation techniques, where the parameters to be estimated are the locations (on the frequency axis) and amplitudes of the spline knots. We choose to use Akima splines~\cite{akima1970} to represent the log-PSD of each noise component $\alpha$, for which the parameters are the locations and amplitudes of the knot points $x_i = \log f_i, y_i = \log S_i$. For any log-frequency $x = \log f$ between $x_i$ and $x_{i+1}$, we have 
\begin{eqnarray}
    \log S_{\alpha}(x) = \sum_{p=0}^{3} a^{\alpha}_{i, p} (x - x_i)^{p}, \; \forall x \in [x_i, \, x_{i+1}]
\end{eqnarray}
where 
\begin{align}
    a_{i, 0} & = y_{i} \; ;\nonumber \\
    a_{i, 1} & = \frac{m_{i-1} + m_{i}}{2} , \; m_{i} = \frac{y_{i+1} - y_{i}}{x_{i+1} - x_{i}} \; ;\nonumber \\
    a_{i, 2} & = \frac {3m_{i}-2s_{i}-s_{i+1}}{x_{i+1}-x_{i}} \; ;\nonumber \\
    a_{i, 3} & = \frac {s_{i}+s_{i+1}-2m_{i}}{(x_{i+1}-x_{i})^{2}}.
\end{align}
The Akima splines are considered to be very stable locally, and therefore do not exhibit Gibbs-like ringing or overshoot, in contrast to the standard cubic B-spline interpolation functions, which often resulted in spurious \gls{psd} estimates for close-by spline knots. In turn, spurious \gls{psd} curves may cause numerical issues in the likelihood calculation. The above are reasonable justifications to move away from the spline functions used originally in~\cite{Baghi:2023qnq}. The Akima splines were also adopted recently in~\cite{Gupta:2023jrn} for similar reasons.

One of the common challenges with shape-agnostic models is the actual number and placement of spline knots along the $x$ axis. Depending on the measured data, too many spline knots might result in overfitting issues, while too few might fail to capture the data complexity. As already described in previous subsection~\ref{sec:modelselect}, we solve this problem by adopting trans-dimensional sampling techniques, where the number of knots $k$, their positioning in log-frequency space $x_{j,k} = \log f_{j,k}$, and their amplitudes $\log S_{j,k}$ are determined simultaneously from the data. We remind the reader here that the $j$ index corresponds to the spline number for the given model order $k$.

In~\cite{Baghi:2023qnq}, the optimal model order for such a configuration was found to be $k=6$. This configuration was suitable for the model assumptions of~\cite{Baghi:2023qnq}. In the present work, \gls{tm} and \gls{oms} noises are reflected with different transfer functions (as described in section~\ref{sec:modelselect}), and their interferometric-level \gls{psd} $S_{\alpha}$ have distinct shapes. Therefore, a full spectral-shape agnostic model, for the case of equal noise \glspl{psd} across interferometers, requires three dynamical spline-interpolation models: two for the overall instrument noise and one for the \gls{sgwb} signal. To decrease the number of spline knots necessary to characterize the data, we fit deviations from reference log-spectra (taken equal to the injection). This practice restrains the dimensionality of the problem. We have also repeated the analysis on the actual spectra rather than the deviations from the injected ones to verify that we obtain similar results. The difference lies in the much larger required dimensionality of the model, which translates to longer \gls{mcmc} runs to achieve convergence and therefore an increase of required computational resources.

\subsection{Parametric spectral models}
\label{sec:template-models}
To investigate the comparative use of parametric templates for signal and noise in \gls{sgwb} searches, we consider the analytical models described below.

\subsubsection{Noise template}
\label{sec:noise-template}

The \gls{psd} template of each noise component depends on one amplitude parameter $a_{\alpha}$, with $\alpha = \{\mathrm{TM}, \mathrm{OMS}\}$.
The \gls{tm} noise \gls{psd} is
\begin{align}
	S_{\mathrm{TM}}(f) = a_{\mathrm{TM}}^2 \left[1 + \left(\frac{f_{1}}{f}\right)^2 \right] \left[1 + \left(\frac{f}{f_2}\right)^4\right],
\label{eq:tm}
\end{align}
where $f_{1} = 4 \times 10^{-4}$ Hz and $f_2 = 8$ mHz. The same model is used for all data injections with the value $a_{\mathrm{TM}} = 3 \times 10^{-15}\,  \mathrm{m s^{-2}}$.
The readout noise \gls{psd} is
\begin{align}
	S_{\mathrm{OMS}}(f)= a_{\mathrm{OMS}}^2 \left[1 + \left(\frac{f_3}{f}\right)^4\right],
\label{eq:oms}
\end{align}
where $f_3 = 2$~mHz. In the data injection, we set $a_{\mathrm{OMS}} = 15 \times 10^{-12} \, \mathrm{m Hz^{-1/2}}$.

\subsubsection{Signal template}
\label{sec:signal-template}

The template for the \gls{sgwb} is an analytic power-law model, which is also used for the data injections. This model is written as 
\begin{equation}
	S_{\mathrm{GW}}(f) = \Omega_\mathrm{GW} \left(\frac{f}{f_0}\right)^{n} \frac{3H_{0}^2}{4 \pi^2 f^3},
	\label{eq:sgwb_psd}
\end{equation}
where the parameter set $\boldsymbol{\theta}_{\mathrm{GW}}\equiv\{ \Omega_\mathrm{GW}, \, n\}$ contains the amplitude and spectral index of the power law, respectively, $f$ is the frequency, and $H_{0}$ is the Hubble parameter at the present day. The frequency $f_0$ is a convention parameter (the so-called pivot frequency) that we set to $f_0=0.003~\mathrm{Hz}$. 

\subsection{Different priors on the instrumental noise}
\label{sec:priors}

\begin{figure}
    \centering
    \includegraphics[width=0.38\linewidth]{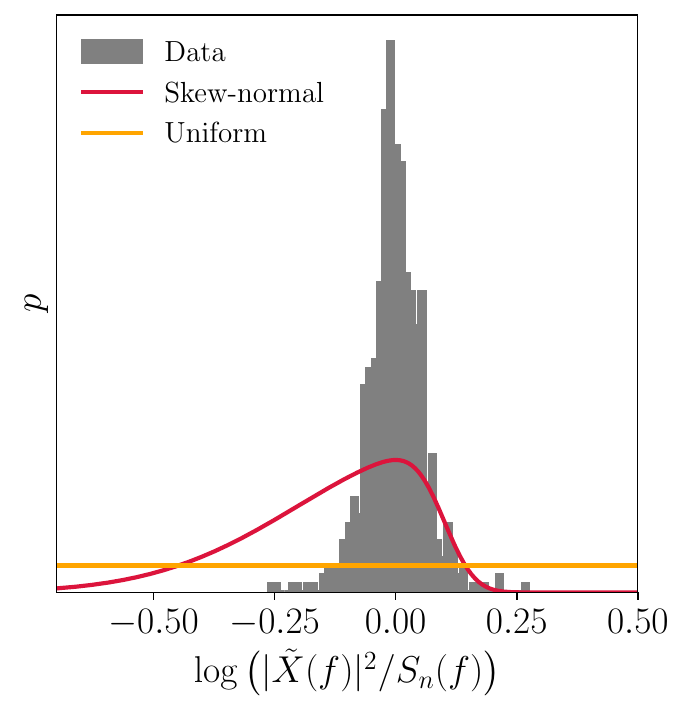}
    \caption{The different priors adopted in our analyses, compared to the histogram of the whitened averaged periodogram of the time-series data (gray). Data taken from~\cite{lisa-sim-data}.}
    \label{fig:data_vs_prior}
\end{figure}

In this work, we investigate the impact of the noise parameter prior probability distributions on the \gls{sgwb} search. We compare two types of priors. The first one is uninformative: it is a wide uniform distribution centered on the true injected noise spectrum, which is chosen as a baseline for comparison with our earlier work~\cite{Baghi:2023qnq}. This means that, for the case of the flexible shape-agnostic spline model, the knot amplitudes are allowed only within the bounds of $\log_{10} S_{j,k} \sim \mathcal{U} [-1,\, 1]$. We also repeat the analysis with an informative prior, which is based on the skewed normal distribution~\cite{2009arXiv0911.2093A}, again centered around the true injected noise spectra. The skewed normal distribution for a random variable $x$ is defined as
\begin{equation}
	\mathcal{SN}(x; \mu, \sigma, \alpha) = \frac{2}{\sigma} \mathcal{N}( x; \mu, \sigma)\Phi\left(x;\frac{\alpha(x-\mu)}{\sigma}\right),
	\label{eq:skewn}
\end{equation}
where $\mathcal{N}$ is the normal probability density function and $\Phi$ its cumulative distribution function. The parameters $\mu,\,\sigma$ are the distribution location and scale, respectively, while $\alpha$ is an overall shape parameter. For the analysis with informative priors, we have chosen $\alpha=-6$, $\mu=0.1$, and $\sigma=0.3$, which result in a distribution with its mode around the injected noise \gls{psd}. A histogram of the data \gls{psd} versus the different priors assumed is shown in figure~\ref{fig:data_vs_prior}. 
This choice of asymmetric prior is motivated by the fact that if the instrumental performance meets the mission requirements, the actual noise level has a larger probability to lie below the requirement curve than above.
Besides the fully agnostic analysis, we also perform a baseline analysis with analytic models for both the noise and the signal. The noise models are those of eq.~(\ref{eq:tm}) and (\ref{eq:oms}). We set priors on the deviations from the injected ``spectral amplitude''  parameters of $\log_{10}a_{\mathrm{TM}}$ and $\log_{10}a_{\mathrm{OMS}}$, both allowed within the bounds of $\mathcal{U} [-1,\, 1]$, or follow the skewed normal of eq.~(\ref{eq:skewn}) with the same $\alpha, \mu, \sigma$ parameters mentioned above. 

\subsection{Priors on the stochastic gravitational wave signal}
\label{sec:signalpriors}

\begin{figure}
    \centering
    \includegraphics[width=1.0\linewidth]{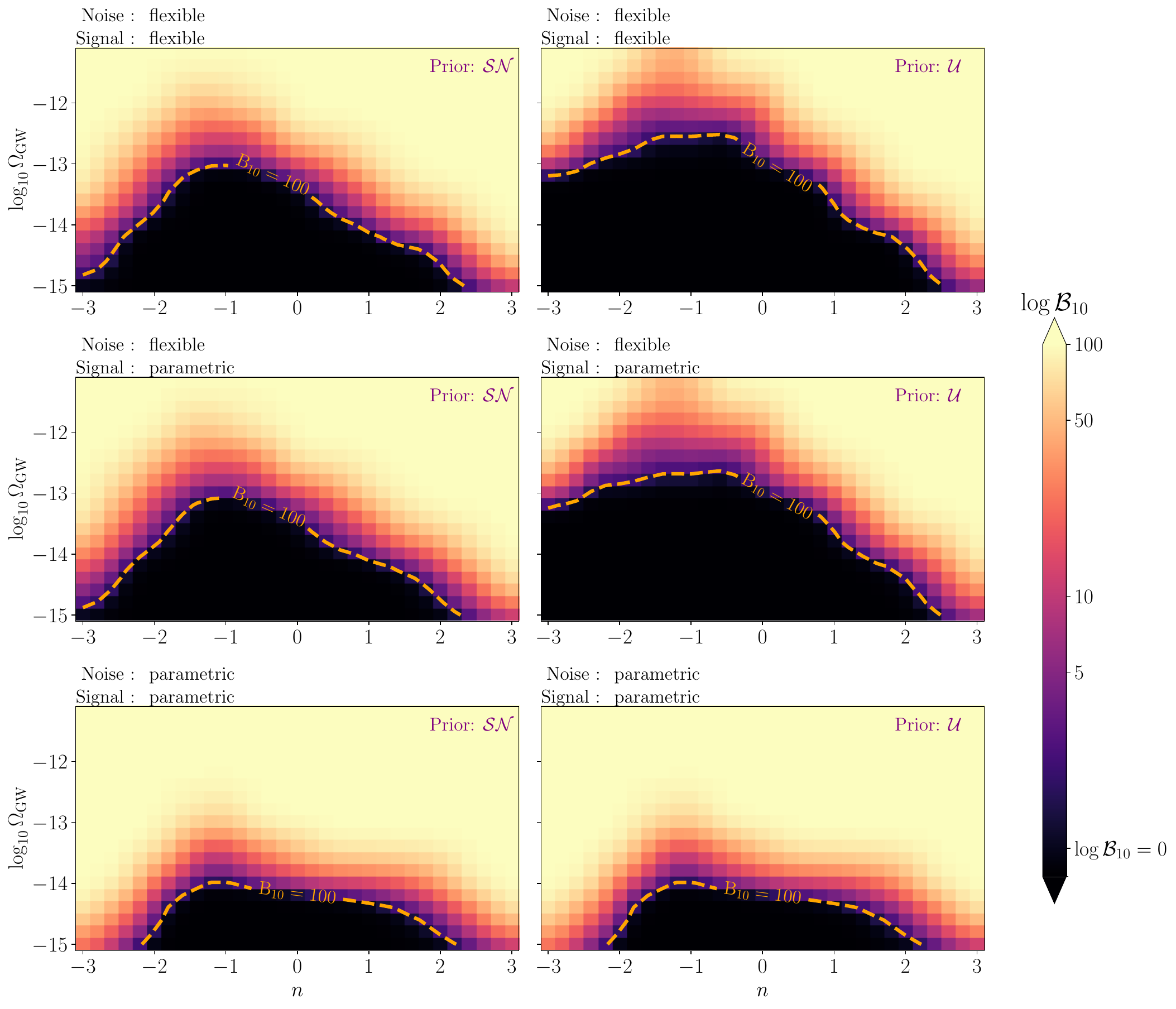}
    \caption{Bayes Factors $\mathcal{B}_{10}$ between the noise-and-signal $H_{1}$ and the noise-only $H_{0}$ hypotheses, for different injections of \gls{sgwb}, under the assumptions of various priors on the noise model, and also different model types (parametric versus flexible). For each panel, the $\log_{10}\Omega_\mathrm{GW}$ and $n$ are the amplitude and spectral index respectively of the power law of the injected signal. The first row contains the results of the runs where flexible models were used for all the noise and signal models, while the second row shows the results where a parametric model was used only for the signal. The last row shows the results obtained when using parametric models for all contributions to the covariance matrix. The first (respectively the second) column shows the results of the runs where the informative $\mathcal{SN}$ (respectively non-informative $\mathcal{U}$) prior was used on the instrumental noise component.}
    \label{fig:bfs}
\end{figure}

As already described, we adopt two types of models for the signal analysis: a template-based and a spline-based description. When using the template model described in \ref{sec:signal-template}, the priors for the \gls{sgwb} signal are rather wide uninformative priors of $\log \Omega_\mathrm{GW} \sim \mathcal{U} [-20,\, -9]$ and $n \sim \mathcal{U} [-5,\, 5]$, in order to cover the complete parameter space that potentially yields detectable signals. When the agnostic spline model (described in \ref{sec:spline-models}) is used to fit the signal, we choose a wide uniform prior for the spline knots, with a maximum allowed power as $\log S_{\mathrm{GW}}(f, \boldsymbol{\theta}_{\mathrm{GW}}) \sim \mathcal{U} [-135,\, -71]$ for all frequencies. We expect improved detectability capabilities for a given \gls{sgwb} signal when the prior assumptions are informative and when the fitting models are templates depending on fewer parameters. We verify this expectation in the sections below. 

\section{Simulations and analysis}
\label{sec:results}

As in~\cite{Baghi:2023qnq}, we choose to work with the expectation values for the data instead of realizations. In particular, if $\mathbf{\bar{Y}} = \operatorname{E}_{\boldsymbol{\theta}^{\star}}[\mathbf{Y}]$ is the expectation of the data $\mathbf{Y}$ under the true hypothesis, then $\overline{\log \mathcal{B}_{10}} = \mathcal{B}_{\mathrm{thresh}}$ yields a conservative criterion for detection, for a given $\mathcal{B}_{\mathrm{thresh}}$. We also choose $\mathcal{B}_{\mathrm{thresh}}=100$ as the threshold for confident signal detection.

We also consider a \gls{lisa} constellation that follows analytic Keplerian orbits~\cite{bayle_2025_orbits}. In this configuration, arm-length variations are not large enough to cause significant non-stationarities in the data. More realistic orbits would require a time-frequency analysis strategy~\cite{Buscicchio:2025zeb}. In addition, we consider the test-mass acceleration and interferometric noises to be constant and equal between spacecraft for the duration of the simulation ($T_\mathrm{obs} = 1~\mathrm{year}$). This is a simplified scenario that reduces the computational requirements of the analysis, because we can use common spectral models for all spacecraft noises simultaneously. A thorough investigation of the impact of asymmetric noise is left for future work.

As a variety of stochastic signals from cosmological or astrophysical origins are predicted to follow a power-law at the $\mathrm{mHz}$ range ~\cite{LISACosmologyWorkingGroup2022kbp}, we adopt this type of model for our injections. Thus, we create a grid of amplitudes and spectral indices $\boldsymbol{\theta}_{\mathrm{GW}}\equiv\{ \Omega_\mathrm{GW}, \, n\}$, essentially covering the parameter space that is potentially covered by the \gls{lisa} sensitivity. The grid of injections spans between $10^{-15} < \Omega_\mathrm{GW} < 5\times 10^{-12}$ and $-3 < n < 3$. For each pair of $\boldsymbol{\theta}_{\mathrm{GW}}$, we run six types of inferences: parametric (analytic) and flexible (spline) models for signal and noise, both with informative and uninformative prior assumptions. For each injection, we also run the noise-only case, intending to compute the Bayes factors and assess the detectability of the given signal under the different modelling assumptions. Injections always contain noise and signal.

Finally, as already mentioned, when using the flexible spline approach, we always infer the optimal number of spline knots using \gls{rj} sampling methods. In the end, for each run we also compute the evidence for the given model using the Stepping Stone algorithm, as introduced in section~\ref{sec:modelselect}. The evidence computation of the full trans-dimensional problem is achievable because only the temperatures and likelihoods for each \gls{rj}-\gls{mcmc} walker are needed~\cite{Karnesis:2023ras}. The noise-only versus the noise-plus-signal hypotheses are then tested by computing the Bayes Factor as the ratio of the evidences calculated. The sampler was configured to run \gls{mcmc} chains for $50$ walkers, for each of the $30$ temperatures used in the adaptive parallel tempering scheme. After a short burn-in period, during sampling computations each second \gls{mcmc} point was kept in the chain until a total $10^4$ \gls{mcmc} samples were gathered for each walker. The results of our analysis are summarized in figure~\ref{fig:bfs} and in the sections below. Example reconstructions of the recovered signals can be seen in figure~\ref{fig:reconstruction}. 

\subsection{Impact of prior and model choice}
\label{sec:prior-impact}

\begin{figure}[!tbp]
  \centering
  \subfloat[]{\includegraphics[width=0.44\textwidth]{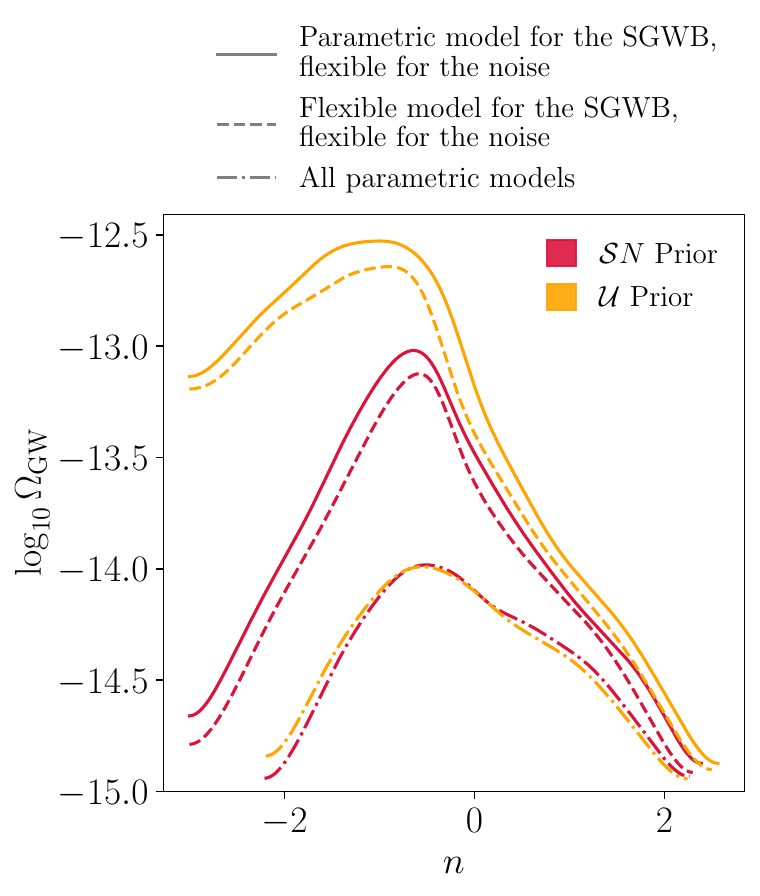}\label{fig:model_impact}}\hspace{1cm}
  \subfloat[]{\includegraphics[width=0.42\linewidth]{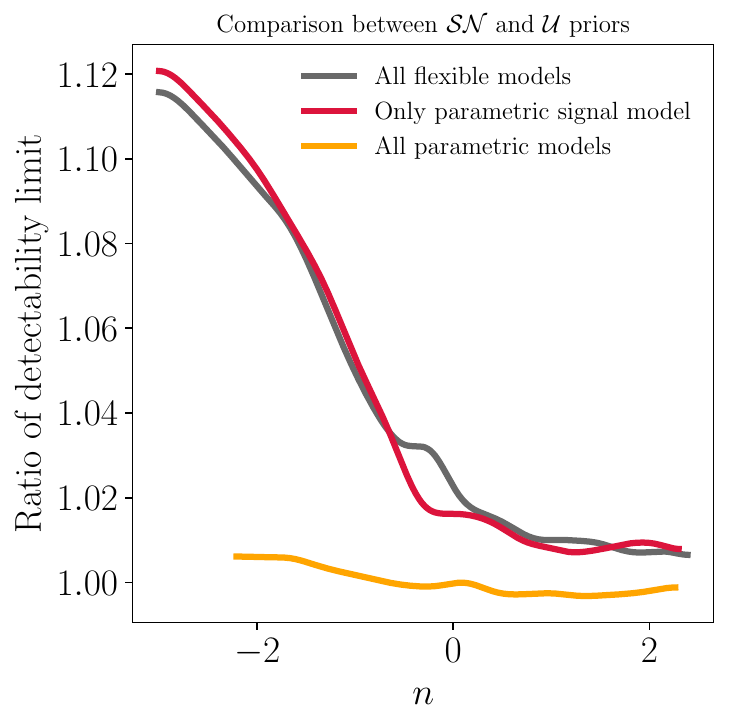}\label{fig:prior_impact}}
  \caption{(a) The detectability limits as functions of the adopted type of models for the noise signal and the prior used on the noise components. The color denotes the type of prior used for the noise \gls{psd}, while the line style denotes the type of model. This panel is basically a summary of figure~\ref{fig:bfs}. (b) Comparison of the detectability limits for different types of models with respect to the assumed prior on the spectral shape of the noise. This figure presents the ratio between the curves of the two column panels of figure~\ref{fig:bfs}, or the ones from the left panel of this figure. The choice of informative prior is more impactful at the lower part of the frequency spectrum, where the noise and signal transfer functions are comparable. We note a similar behavior for both types of \gls{sgwb} model. This does not hold true for the case were analytic models where adopted for all components of the data covariance matrix (orange line). We remind the reader that the $\log_{10}\Omega_\mathrm{GW}$ and $n$ are the amplitude and spectral index respectively of the power law of the injected signal.}
\label{fig:impact}
\end{figure}

In figure~\ref{fig:bfs} we plot the Bayes Factors $\mathcal{B}_{10}$ between the noise-plus-signal $H_{1}$ and the noise-only $H_{0}$ hypotheses for injected power-law signals of varying amplitudes and spectral indices. Each row corresponds to a different analysis model. The first row shows the results for the full flexible models for both signal and noise. In the second row, we show the results using a template for the signal, while the noise description remains flexible. In the third row, we present the results of the analyses using the analytic models for signal based on eq.~(\ref{eq:sgwb_psd}) and noise based on eqs.~(\ref{eq:tm}) and (\ref{eq:oms}). 

Each column corresponds to a different noise prior. The left-hand side column shows results with the informative (skewed-normal) prior, while the right-hand side column shows results with the non-informative (uniform) prior. For all analyses, we have chosen the conservative limit of $\mathcal{B}_{10}=100$ as the threshold for a confident detection of a stochastic signal~\cite{Kass:1995loi}.
\begin{figure*}[t!]
        \subfloat[\textit{(Left)}: Flexible models for both the signal and the noise, with a skewed-normal prior on the noise spectral shape. \textit{(Right)}: Same as in left panel, but with a broad uniform prior on the noise.]{%
            \includegraphics[width=.44\linewidth]{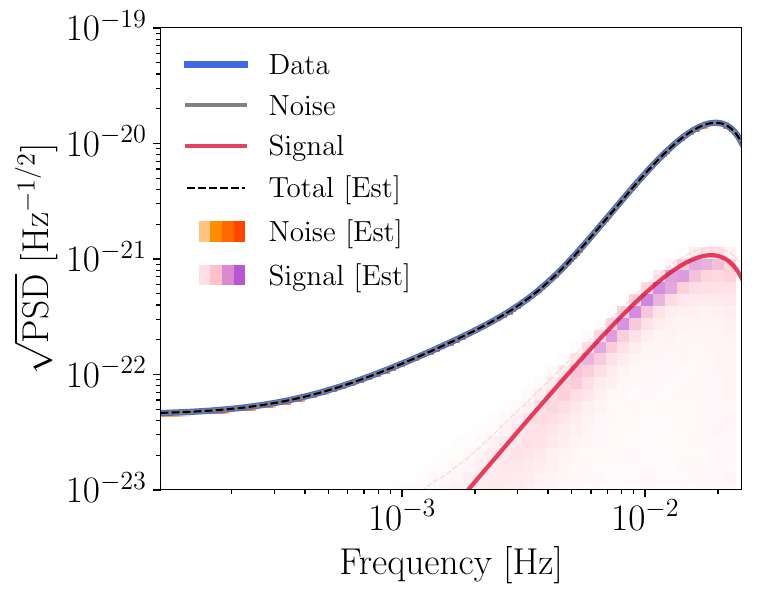}\hspace{.7cm}%
            \includegraphics[width=.44\linewidth]{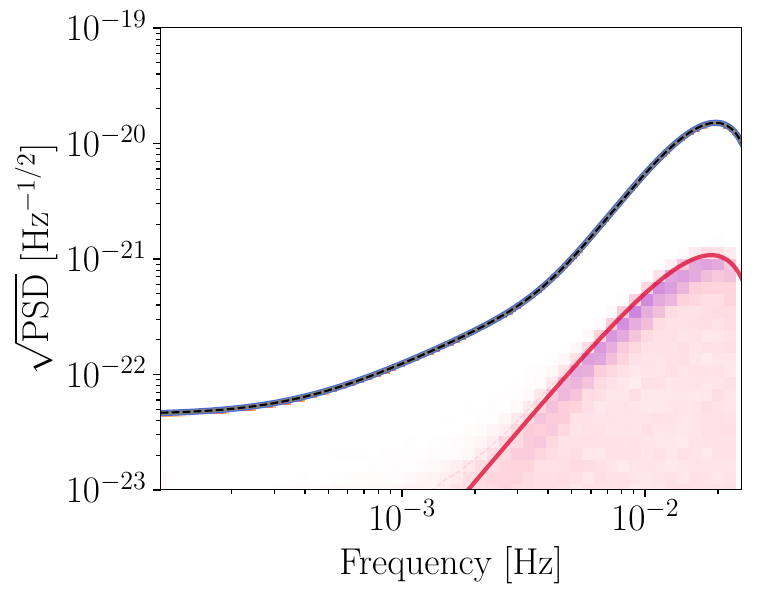}%
            \label{fig:reconstruction-all-splines}}
     \\
     \subfloat[\textit{(Left)}: Flexible model for the noise, parametric model for the \gls{sgwb} signal, using the skewed-normal prior on the noise. \textit{(Right)}: Same as in left panel, but with a uniform prior on the noise.]{%
             \includegraphics[width=.44\linewidth]{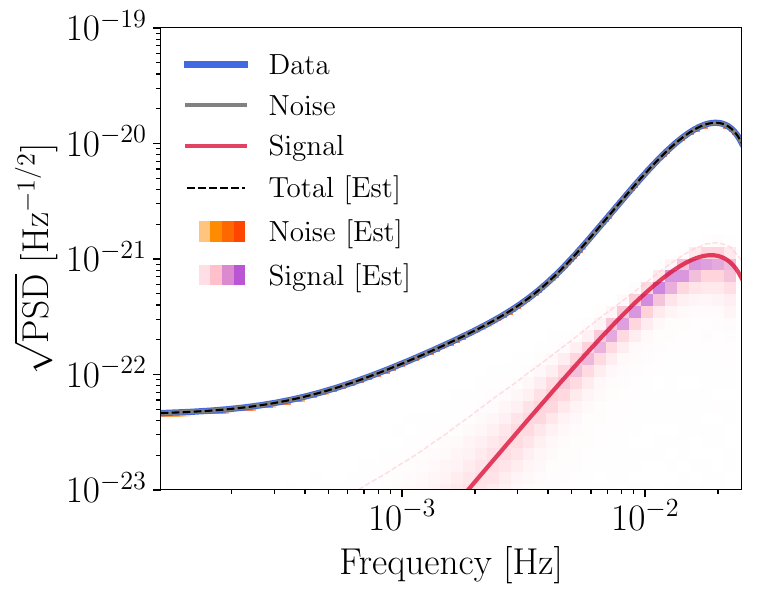}\hspace{.7cm}%
             \includegraphics[width=.44\linewidth]{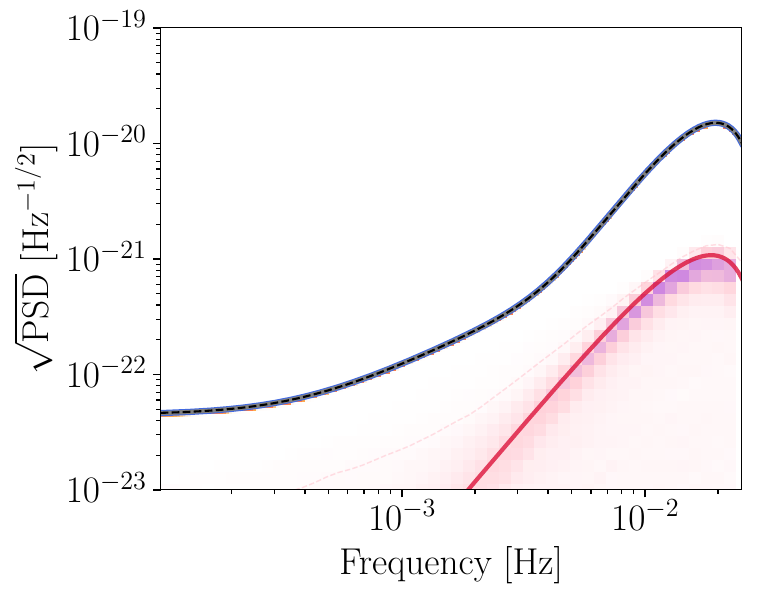}
             \label{fig:reconstruction-n-spline-s-analytic}}
    \\
     \subfloat[\textit{(Left)}: Parametric models for both the \gls{sgwb} signal and the noise, and skewed normal prior on the noise. \textit{(Right)}: Same as in left panel, but with a uniform prior on the noise.]{%
             \includegraphics[width=.44\linewidth]{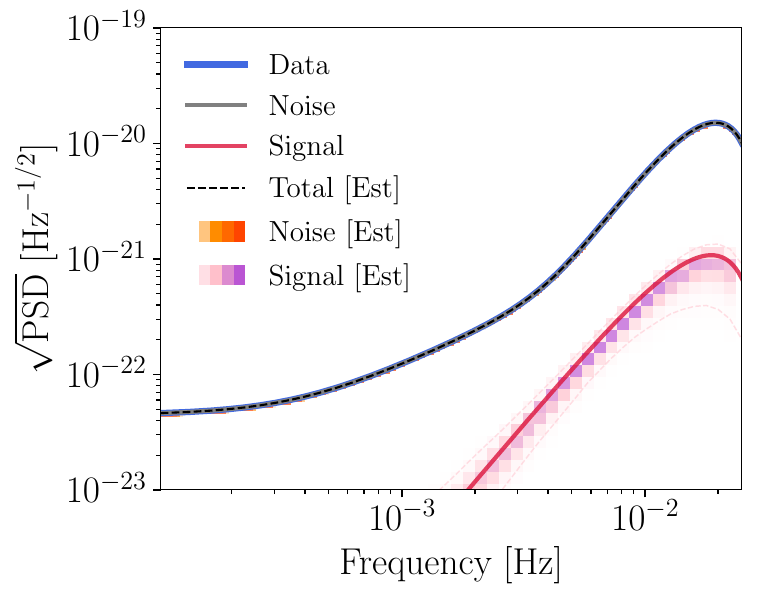}\hspace{.7cm}%
             \includegraphics[width=.44\linewidth]{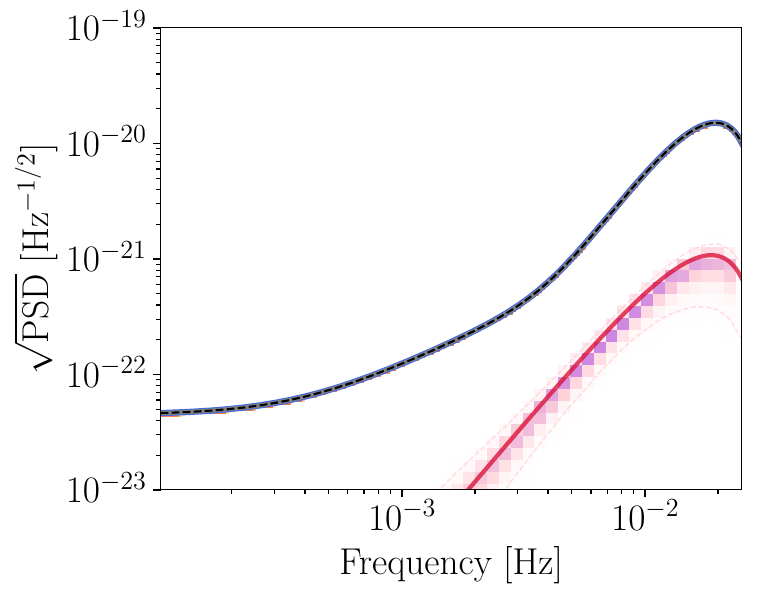}%
             \label{fig:reconstruction-n-a-s-a}}%
        \caption{Examples of signal reconstruction for an injection of a \gls{sgwb} that follows the shape of a power-law with $\Omega_\mathrm{GW}=5\times10^{-15},\, n=2$. The analysis has been performed under different assumptions about the noise priors and \gls{sgwb} spectral models.}
        \label{fig:reconstruction}
\end{figure*}
The first row demonstrates that the prior has a significant impact on the detectability of a \gls{sgwb}. The more informative skewed-normal prior yields larger Bayes factors, which in turn pushes the detectability bounds to weaker signals. This holds for any type of model assumed for the signal and the noise. This is evident from figure~\ref{fig:prior_impact}, where we compare the detectability bounds for the same injections under the two different prior assumptions ($\mathcal U$ versus $\mathcal SN$). The curves of figure~\ref{fig:prior_impact} are the ratio of the dashed lines between the column panels of figure~\ref{fig:bfs}. The informative prior has more impact at lower frequencies, where the transfer function of the test-masses acceleration noise, which is the dominant noise component, is similar to the transfer function of the \gls{gw} signal. The use of a non-informative prior greatly reduces the detectability capabilities of LISA, except in the case where analytic models were used for all components of the data covariance matrix (see bottom row of figure~\ref{fig:bfs} and orange curve of figure~\ref{fig:prior_impact}).
\begin{figure}[!tbp]
  \centering
  \subfloat[]{\includegraphics[width=0.6\textwidth]{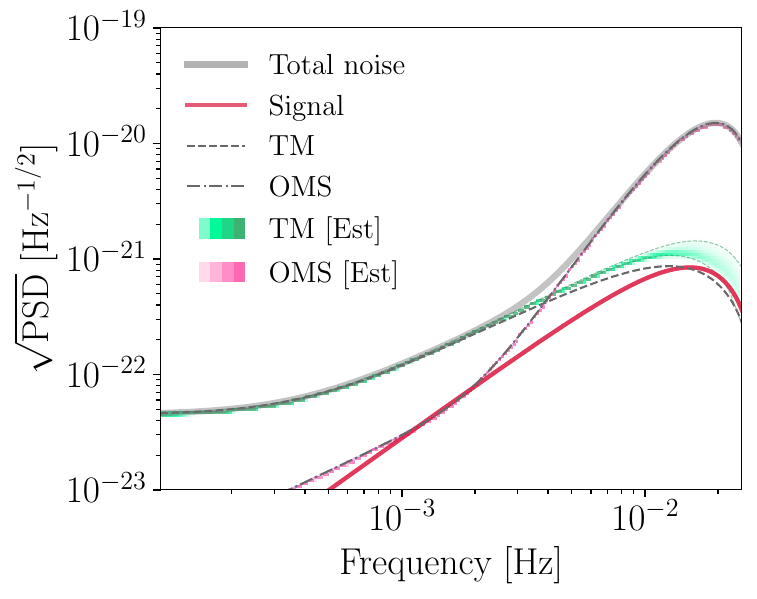}
  \label{fig:sig-absorb-spectrum}}
  \subfloat[]{\includegraphics[width=0.4\linewidth]{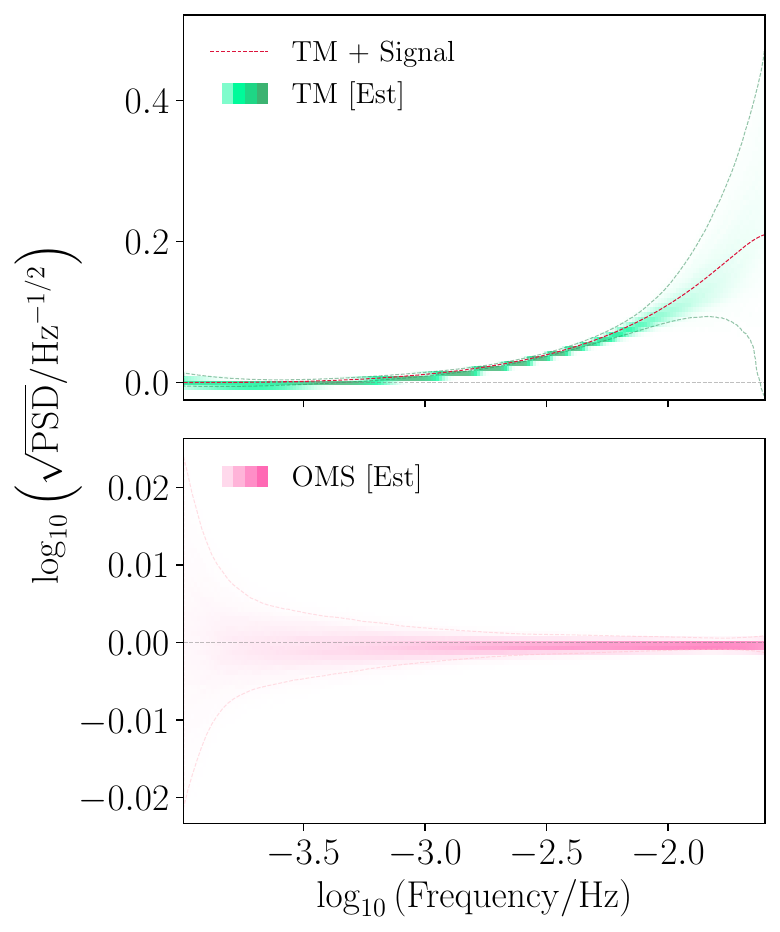}\label{fig:signal-absorb-residuals}}
  \caption{Demonstration of a case where the power induced by a relatively weak isotropic \gls{sgwb} signal can be fully absorbed by the flexible noise models. For this investigation we inject a \gls{sgwb} signal with $\Omega_\mathrm{GW}=10^{-13},\, n=0$, while we fit for only the noise using flexible spline models. (a) The reconstructed noises of the \gls{tdi} X channel are shown with the green heatmap for the \gls{tm} and with the pink for the \gls{oms} noises respectively. The injected true spectra are shown with the dashed and dot-dashed gray lines respectively, while the injected \gls{sgwb} signal with red. (b) A zoom in on the deviations from the injected noise spectra. The gray dotted line represents the baseline around zero. The dashed red line represents the sum of the true injected spectra of \gls{tm} and \gls{sgwb} signal. It is evident that the power induced by the \gls{sgwb} signal is absorbed by the \gls{tm} noise model.}
\label{fig:signal-absorbtion}
\end{figure}
Although the choice of model type for the signal (parametric and flexible) brings an improvement, it has a smaller effect on the detectability bounds on the parameter space than the choice of prior. In figure~\ref{fig:model_impact} we present a direct comparison of the impact of the different models used in the analysis, using different line styles. As expected, a data-consistent parametric template for the signal enables the detection of a \gls{sgwb} in a larger portion of the parameter space when compared to the flexible case. This can be attributed to the lower dimensionality of the model, which simultaneously controls and constrains the different stochastic contributions across the frequency range. When using a parametric (analytic) model for both the noise and the signal, this advantage diminishes, as shown by the overlapping dashed-dotted lines of figure~\ref{fig:model_impact}. In figure~\ref{fig:prior_impact}, we show the ratio between the detectability bounds assuming different priors for the \gls{psd} of the noise. It is the ratio between the solid lines of different colors of figure~\ref{fig:model_impact}. Examples of the reconstruction of the injected stochastic signal are given in figure~\ref{fig:reconstruction}, while more information can be found in the Appendix~\ref{sec:complementary-figures}, and in its figures~\ref{fig:prior_impact-analytic-sgwb}, where we compare the parameter estimates of the \gls{sgwb} signal with a power-law model under different prior assumptions. In figure~\ref{fig:knots_k}, we show an example of the marginal posterior of the spline knots amplitude and position on the frequency axis for the two noise components of the instrument. 

\subsection{Flexible models detectability limit and impact of prior range}
\label{sec:failures}

Flexible spectral models may introduce degeneracies in the analysis. For example, we have found that a marginally detectable isotropic stochastic signal can be misinterpreted as excessive power of the \gls{tm} noise. This heavily depends on the choice of noise prior, as well as the type of model for the stochastic signal. Such a situation is presented in \cref{fig:signal-absorbtion}: we have injected a marginally detectable signal with $\Omega_\mathrm{GW}=10^{-13}$ and $n=0$ (see \cref{fig:bfs}), and then analyze the data with flexible spline models for the noise and the signal. Here, we only fit for the noise. In the left panel~\ref{fig:sig-absorb-spectrum}, we show the noise budget as estimated with the flexible noise models; in the right panel~\ref{fig:signal-absorb-residuals}, we focus on the deviations of the estimated spectra from the injected true noise. It is clear that a fraction of the signal power has been absorbed by the model that aims to characterize the \gls{tm} noise contribution. This effect is naturally more prevalent for weak stochastic signals, and for uninformative uniform priors for the noise power spectra.

\begin{figure*}[t!]
        \includegraphics[width=.44\linewidth]{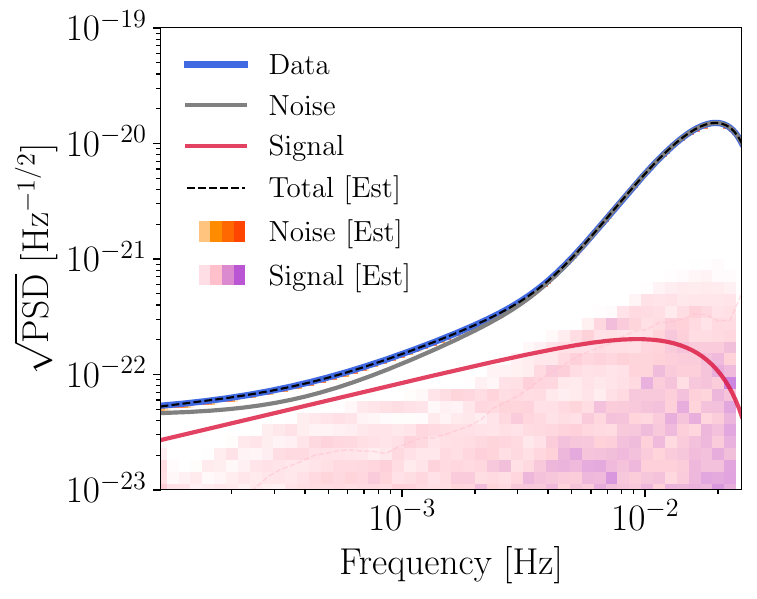}\hspace{.7cm}%
       \includegraphics[width=.44\linewidth]{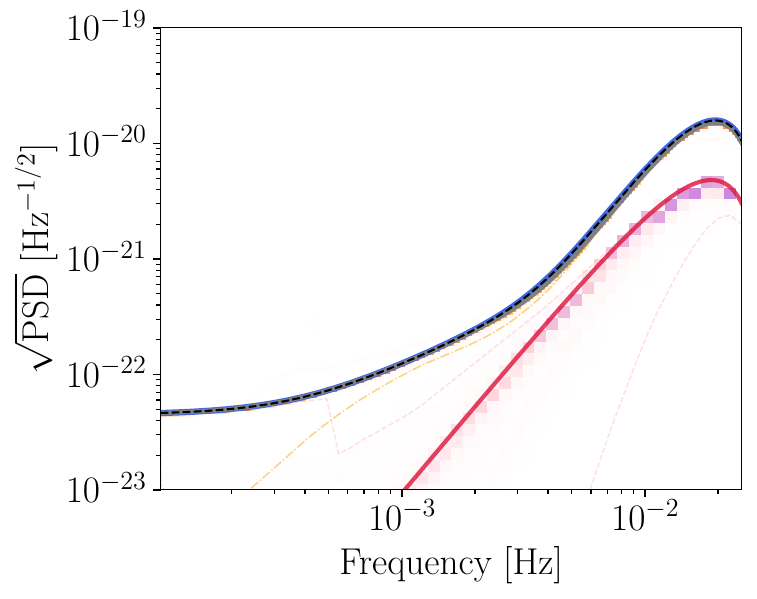}%
             \label{fig:reconstruction-n-a-s-a}
        \caption{Analysis results of two extreme scenarios with very broad uniform priors on the instrumental noise. The two injected signals have the same amplitude $\Omega_\mathrm{GW}=10^{-13}$, but different spectral indices $n$ (left panel $n=-2$, right panel $n=2$). For both cases, the signal and the noise are fitted with flexible weakly parametric models. See text for details.}
        \label{fig:reconstruction-fail}
\end{figure*}

One can consider even more extreme scenarios where the prior on the noise is much broader than in our previous analyses. To study the performance of our pipeline in such extreme cases, we have simulated two \gls{sgwb} signals with the same amplitude $\Omega_\mathrm{GW}=10^{-13}$, but with different spectral indices ($n=-2$ and $n=2$, respectively). For these tests we adopt flexible spline models for both signal and the noise. Both injected signals, according to \cref{fig:bfs}, were confidently detectable given the choice of priors in the analysis (see \cref{sec:results}). \Cref{fig:reconstruction-fail} shows the results of the analysis when adopting the very wide uniform priors $\log S_{j,k} \sim \mathcal{U} [-10,\, 10]$ (note the natural logarithm here) for the spline knots amplitudes. In the left panel of \cref{fig:reconstruction-fail}, it is evident that, by adopting very broad uniform priors, the injected signal with $n=-2$ is undetectable ($\log \mathcal{B}_{10} = -33.8\pm 0.3$). This is an expected outcome, which can be attributed to a combination of prior and the similarities of the transfer functions of the \gls{sgwb} and the \gls{tm} noise, where the latter is dominant at the lower part of the spectrum (see \cref{fig:sig-absorb-spectrum}). However, as shown in the right panel of the same figure, the signal with $n=2$ remains detectable, albeit with decreased confidence than the previous analysis with tighter uniform priors ($\log \mathcal{B}_{10} = 23.8 \pm 0.8$).  

We can conclude that the choice of prior on the noise has a significant impact on the search of stochastic signals, and extreme cases of wide uninformative priors may cause degeneracies in the analysis. In reality, we expect that the prior on the noise will be physically motivated from instrument characterization experiments and from auxiliary channels measuring the environmental conditions inside the satellites. At the same time, we have also shown that one can exploit the differences of the transfer functions of the noise components in order to improve their resolvability. However, we should point out that given the above results, exploiting the differences of the transfer functions is more effective in the high-frequency part of the spectrum of the data. For the case of flexible spline models, the Bayesian Occam's Razor penalty induced by the increased dimensionality of the model (more spline knots needed) could also help discriminate between stochastic signals and the noise, since we would expect that a \gls{rj}-\gls{mcmc} algorithm would converge to higher dimensional spline models to correct for small deviations between the different transfer functions of the components of the covariance matrix. This possibility however, remains to be further studied in future work.


\section{Discussion}
\label{sec:discussion}

We have implemented a data analysis pipeline focusing on the search and parameter estimation of \gls{sgwb} signals in the future \gls{lisa} data. With this framework, we model the noise and potential stochastic signal on the link level of the constellation and then we combine the data to build the final \gls{tdi} measurements to be used in the inference. The transfer functions of the different stochastic components are derived from the orbits of the constellation, and therefore are considered known in the analysis. We use different prior assumptions on the instrumental noise, as well as different types of models, in order to assess the detectability of \gls{sgwb} signals of given characteristics. The analysis presented here is based on our initial work of~\cite{Baghi:2023qnq}.

The first type of models used is a fully parametric model (with $\boldsymbol{\theta}_{\mathrm{GW}}\equiv\{ \Omega_\mathrm{GW}, n\}$ for the signal and $\boldsymbol{\theta}_{\mathrm{n}}\equiv\{ a_{\mathrm{TM}}, a_{\mathrm{OMS}}\}$ for the instrument noise), for which the spectral shape is considered known apriori. The second type is the more flexible model based on spline interpolation methods. This type of model is able to capture unmodeled features on the spectral shape of the data, allowing for the search and reconstruction of virtually any \gls{sgwb} signal or instrumental noise. The optimal model order is inferred from the data with the use of trans-dimensional sampling methods (\gls{rj}-\gls{mcmc} and the {\tt Eryn} package~\cite{Karnesis:2023ras}). In both cases, we model the different components of the data covariance matrix at interferometer level, and apply the known transformation matrices to form the \gls{tdi} combinations. This safeguards against spurious effects caused by the imperfect modelling of the \gls{tdi} spectrum around the areas where it falls to zero. At the same time, we avoid modelling complicated features of the spectra, which directly translates to simpler models with smaller dimensionality. 

We have considered a series of investigations where we injected power-law \gls{sgwb} signals at different amplitudes and spectral indices. We have then performed the inference for both noise-only $H_0$ and signal-plus-noise $H_1$ hypotheses, which were then used to compute the Bayes factor $\mathcal{B}_{10}$ between the two. This analysis was repeated for the different model types, as well as for different priors on the noise spectra. The informative prior used is based on the skewed-normal distribution, which was centered around the true noise spectra (see figure~\ref{fig:data_vs_prior}), simulating the optimal situation where the noise budget of the instrument is measured and well understood. We also used an uninformative prior based on a wide uniform distribution. The results of this investigation are shown in figure~\ref{fig:bfs}, where the detectability bounds on the parameter space can be drawn. 

As expected, an informative prior knowledge of the instrumental noise greatly affects the detectability of the underlying \gls{sgwb} signal, allowing for a greater portion of the parameter space to be measured by \gls{lisa}. Knowing the spectral shape of the noise brings an improvement, as evident from the lower two panels of figure~\ref{fig:bfs}. A parametric model of the noise allows for improved resolvability of the underlying signal, as opposed to the more flexible models based on spline interpolation methods. On the other hand, the choice of model type of the \gls{sgwb} signal (parametric vs. flexible) does not have as much impact as the prior choice.
We have also tested extreme scenarios with very wide uninformative priors (see section~\ref{sec:failures}) which indicate that \gls{lisa} can potentially detect \gls{sgwb} signals, at least at the high-frequency part of the spectrum.

The above results are expected to vary under more realistic conditions, e.g., with unequal noises across the different spacecraft of the constellation. Similarly, a time-frequency analysis must also be adopted for more realistic orbits. Indeed, in this work, we use simplified analytic Keplerian orbits, which allow us to assume a constant transfer function for the duration of the measurement across the frequency range.

Note that we also have reported Bayes factors computed from a single frequency-domain realization of the instrument noise and \gls{sgwb} signal, with spectra corresponding to their \glspl{psd}. This should constitute a good estimate of the expectation value of the Bayes factors, taken over the noise and signal realizations. Finally, we have assumed conservative uncertainties in the spectral estimates, as we adopt a lower bound for the effective number of degrees of freedom in the Wishart likelihood. This can be refined in the future by accurately including the effect of windowing.

The pipeline presented in this work is versatile and can fit various types of models while adapting their dimensionality. In addition, more recent version of the software can handle time-varying stochastic components, making it suitable for inferring both the non-stationary properties of the instrumental noise as well as anisotropic \gls{sgwb} signals. A prime example is the stochastic signal generated by the unresolved Compact Galactic Binaries. We believe that such a tool will be useful for the analysis of the future data of the \gls{lisa} mission.

\section*{Acknowledgments}
\label{sec:Acknowledgments}

We acknowledge useful and inspiring discussions with F. Pozzoli, R. Buscicchio, M. Muratore, N. Dam Quang, O. Hartwig, M pieroni, A. Santini, R. Meyer, S. Babak, C. Caprini, N. Cornish, H. Inchausp\'e, A. Petiteau, M. Besan\c{c}on, and G. Nardini and the members of the \gls{lisa} \gls{ddpc}. NK was supported by the Hellenic Foundation for Research and Innovation (H.F.R.I.) under the 4th Call for HFRI Research Projects to support Post-doctoral Researchers (Project Number: 28418). 

\section*{Software}
\label{sec:software}

For the analysis and modelling of the \gls{lisa} constellation, we have used the {\tt backgrounds} software\footnote{The package is hosted on a private instance of Gitlab, at \url{https://gitlab.in2p3.fr/qbaghi/backgrounds}.}. We have also utilized the {\tt CudAkima} package by A. Santini~\cite{cudakima_2024_13919394}, and the \texttt{Eryn} sampler for its \gls{rj}-\gls{mcmc} capabilities\footnote{Publicly available at \url{https://github.com/mikekatz04/Eryn}.}. 

\appendix

\section{Complementary figures for GW parameter posteriors}
\label{sec:complementary-figures}

\counterwithin{figure}{section}
\renewcommand\thefigure{\thesection\arabic{figure}}
\setcounter{figure}{0}

\begin{figure}[!tbp]
  \centering
  \subfloat[]{\includegraphics[width=0.4\textwidth]{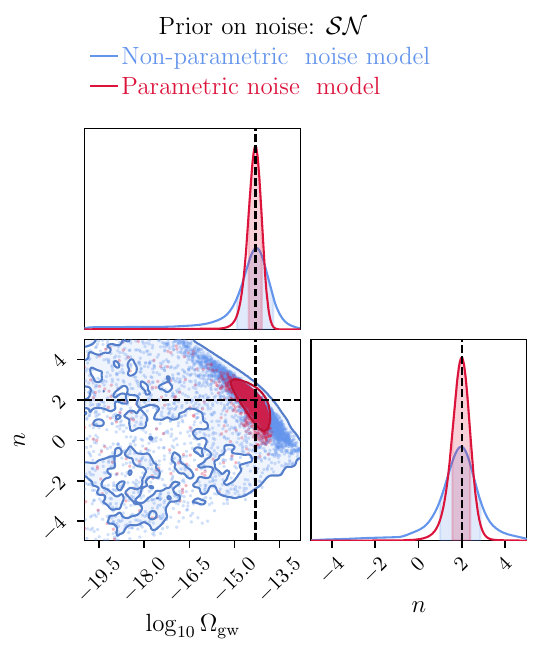}\label{fig:prior_impact-analytic-skew}}\hspace{1cm}
  \subfloat[]{\includegraphics[width=0.4\linewidth]{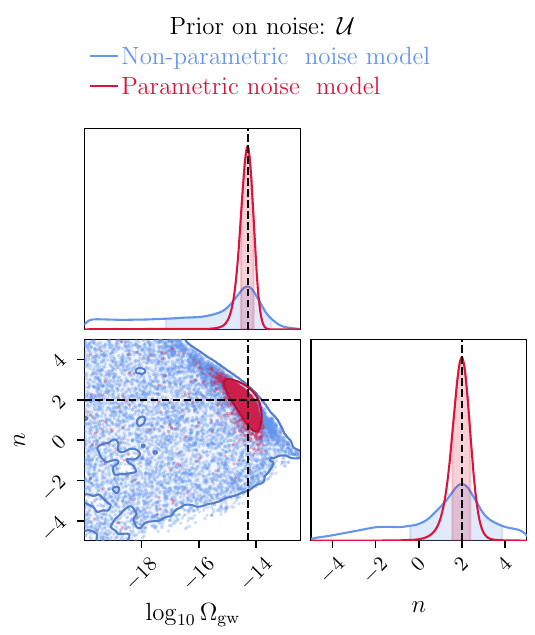}\label{fig:prior_impact-analytic-flat}}
  \caption{ Comparison of the parameter estimates of the \gls{sgwb} signal ($\Omega_0=5\times10^{-15},\, n=2$) with a power-law model, under different prior assumptions, and with different models for the noise (parametric and flexible, based on spline interpolation). (a) Using the informative prior based on the $\mathcal{SN}$ distribution, and (b) using the flat uninformative prior. See text for details. }
\label{fig:prior_impact-analytic-sgwb}
\end{figure}

In this section, we present complementary figures that are relevant to the analysis performed for the production of figure~\ref{fig:bfs}. As already mentioned in section~\ref{sec:inference}, we perform parameter estimation analyses for different injections of power-law \gls{sgwb} signals, under different prior assumptions on the instrumental noise, while using analytic models and flexible models for the signal and noise components of the data. As an example, in figure~\ref{fig:prior_impact-analytic-sgwb}, we show the parameter estimation results for a stochastic signal with $\Omega_0=5\times10^{-15}$ and $n=2$. By the result shown in figure~\ref{fig:bfs}, we should expect this signal to be detectable with very high confidence when using analytic models for all components (red data of figure~\ref{fig:prior_impact-analytic-sgwb}), while the detectability should drop when using the more flexible spline models for the \gls{psd} of the noise components (blue data of same figure). The two panels of figure~\ref{fig:prior_impact-analytic-sgwb} demonstrate the effect of the prior choice. 

\begin{figure}[!tbp]
  \centering
  \subfloat[]{\includegraphics[width=0.4\textwidth]{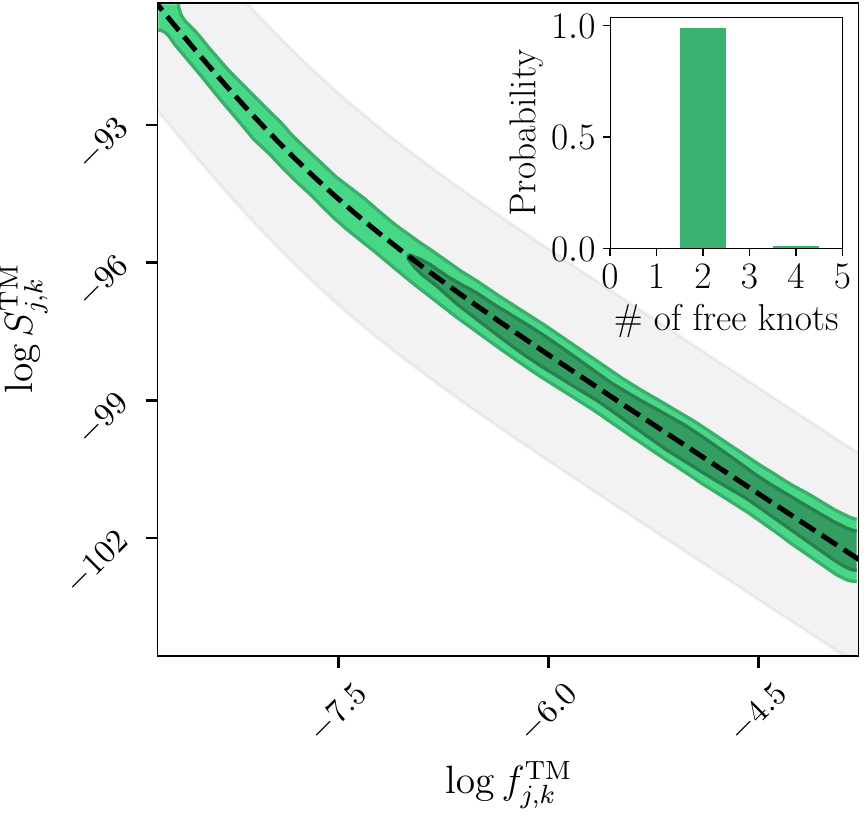}\label{fig:knots-tm}}\hspace{.3cm}
  \subfloat[]{\includegraphics[width=0.4\linewidth]{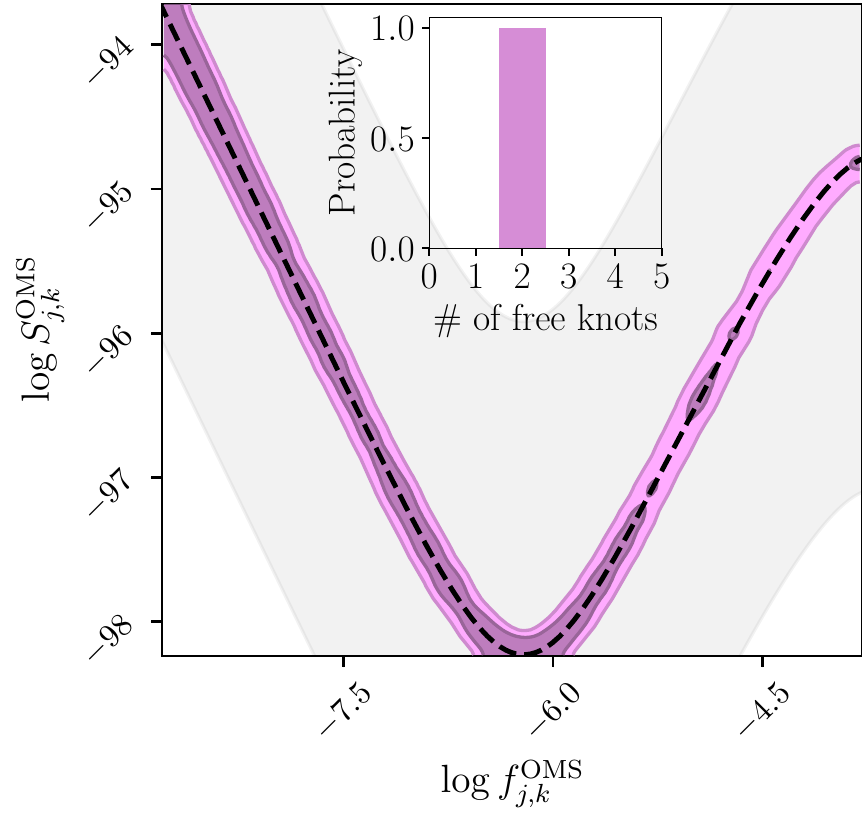}\label{fig:knots-oms}}
  \caption{Examples of the combined posterior of the spline knots amplitudes $\log S_{j,k}$ and position $\log f_{j,k}$ for each $j$ spline number and given model order $k$, for the two instrumental noises, (a) for the \gls{tm} and (b) the \gls{oms} noises. The shaded area represents the uniform prior used for this particular run, while the dashed black line denotes the injected true spectrum. The embedded figure represents the estimated model order, or the number of spline knots, as estimated via  \gls{rj}-\gls{mcmc} sampling. This analysis corresponds to a signal injection with $\Omega_0=5\times10^{-15},\, n=2$ and the final signal estimate shown in figure~\ref{fig:prior_impact-analytic-flat}. See text for more details.}
\label{fig:knots_k}
\end{figure}

In figure~\ref{fig:knots_k}, we show the resulting combined posteriors on the spline knots positions and amplitudes for each model with $k$ knots, as sampled with trans-dimensional \gls{mcmc} methods. The left panel corresponds to the \gls{tm} noise component, while the right panel to the \gls{oms} noise component. We remind the reader that all noises are set to be equal across the three spacecraft, and therefore are being fit simultaneously with the same model in our analysis. The shaded gray area represents the wide uniform prior around the injected signal. Since we fit for deviations from the injected spectra, there is no need for high-order models to capture the complexity of the data, and therefore the number of spline knots estimated converges to the lower prior bound of $k=2$. This is shown in the embedded histograms in both panels of figure~\ref {fig:knots_k}. 

\pagebreak

\providecommand{\noopsort}[1]{}\providecommand{\singleletter}[1]{#1}%

\providecommand{\href}[2]{#2}\begingroup\raggedright\endgroup

\end{document}